\documentclass{aa}  

\usepackage{graphicx}
\usepackage{amsmath}
\usepackage{multirow}
\usepackage{tablefootnote}
\usepackage{txfonts}
\usepackage[colorlinks=true,allcolors=blue]{hyperref}
\defcitealias{1996AJ....112.1487H}{H10}
\newcommand{\fdeg}{\mbox{$. \! ^{\circ}$}}

\begin{document} 

   \title{Precise distances from OGLE-IV member RR Lyrae stars in six bulge
   globular clusters}

   \author{R.~A.~P. Oliveira\inst{1}
          \and
          S. Ortolani\inst{2,3,4}
          \and
          B. Barbuy\inst{1}
          \and
          L.~O. Kerber\inst{5}
          \and
          F.~F.~S. Maia\inst{6}
          \and
          E. Bica\inst{7}
          \and
          S. Cassisi\inst{8,9}
          \and
          S.~O. Souza\inst{1}
          \and
          A.~P\'erez-Villegas\inst{10}
          }

\institute{
Universidade de S\~ao Paulo, IAG, Rua do Mat\~ao 1226, Cidade Universit\'aria,
S\~ao Paulo 05508-900, Brazil;
\href{mailto:rap.oliveira@usp.br}{rap.oliveira@usp.br}
\and
Universit\`a di Padova, Dipartimento di Fisica e Astronomia, Vicolo
dell'Osservatorio 2, I-35122 Padova, Italy
\and
INAF-Osservatorio Astronomico di Padova, Vicolo dell'Osservatorio 5, I-35122
Padova, Italy
\and
Centro di Ateneo di Studi e Attività Spaziali "Giuseppe Colombo" - CISAS. Via
Venezia 15, 35131 Padova, Italy
\and
Universidade Estadual de Santa Cruz, Rodovia Jorge Amado km 16, Ilhéus
45662-000, Brazil
\and
Universidade Federal do Rio de Janeiro, Av. Athos da Silveira, 149, Cidade
Universitária, Rio de Janeiro 21941-909, Brazil
\and
Universidade Federal do Rio Grande do Sul, Departamento de Astronomia, CP 15051,
Porto Alegre 91501-970, Brazil
\and
INAF - Astronomical Observatory of Abruzzo, Via M. Maggini, sn, 64100 Teramo,
Italy
\and
INFN - Sezione di Pisa, Largo Pontecorvo 3, 56127 Pisa, Italy
\and
Instituto de Astronom\'{\i}a, Universidad Nacional Aut\'onoma de M\'exico, A. P. 106, C.P. 22800, Ensenada, B. C., M\'exico
}

\date{}

 
\abstract
   {RR Lyrae stars are useful standard candles allowing one to derive accurate
   distances for old 
   star clusters. Based on the recent catalogues
   from OGLE-IV and 
   \textit{Gaia} Early Data Release 3 (EDR3), the distances can be improved
   for a few bulge globular clusters.}
   {The aim of this work is to derive an accurate distance for the following
   six moderately
   metal-poor, relatively high-reddening bulge globular clusters: NGC\,6266,
   NGC\,6441, NGC\,6626, NGC\,6638, NGC\,6642, and NGC\,6717.}
   {We combined newly available OGLE-IV catalogues of variable stars containing
   mean $I$ magnitudes, with Clement's previous catalogues containing mean $V$
   magnitudes, and with precise proper motions from \emph{Gaia} EDR3.
   Astrometric membership probabilities were computed for each RR Lyrae, in
   order to select those compatible with the cluster proper motions.
   Applying luminosity--metallicity relations derived from BaSTI
   $\alpha$-enhanced models 
   (He-enhanced for NGC\,6441 and canonical He for the other
   clusters),
   we updated the distances with relatively low
   uncertainties.}
   {Distances were derived with the $I$ and $V$ bands, with a $5-8\%$
   precision. We obtained 6.6\,kpc, 13.1\,kpc, 5.6\,kpc, 9.6\,kpc,
   8.2\,kpc, and 7.3\,kpc for NGC\,6266, NGC\,6441, NGC\,6626, NGC\,6638,
   NGC\,6642, and NGC\,6717, respectively. The results are in
   excellent agreement with the literature for all sample clusters, considering
   the uncertainties.}
   {The present method of distance derivation, based on recent data of member
   RR Lyrae stars, updated BaSTI models, and robust statistical methods,
   proved to
   be consistent. A larger sample of clusters will be investigated in a future
   work.} 


\maketitle
%

\section{Introduction}
\label{sec:intro}

Globular clusters (GCs) in the Galactic bulge are important tracers of the
early history of the Galaxy formation \citep[e.g.][]{2018ARA&A..56..223B},
as they keep a memory of the evolution of the Galaxy.
Distances of GCs are the most uncertain information in their studies in terms
of Galaxy structure, stellar population components, and calculation of orbits
\citep[][]{2006A&A...450..105B}. In particular, the orbital parameters of
the clusters projected towards the Milky Way (MW) bulge and their membership
to different Galactic regions are very sensitive to the assumed heliocentric
and Galactocentric distances \citep[e.g.][]{2020MNRAS.491.3251P}. 

These distances and the membership to the bulge region are also of great
interest in the study of high-energy sources since closer distances
imply higher stellar densities and possibly a greater occurrence of such
sources \citep{2007A&A...470.1043O}.
Several bulge GCs host a significant number of X-ray sources and millisecond
pulsars. The most interesting case is Terzan\,5, which contains 39 pulsars
\citep{2005Sci...307..892R, 2018ApJ...855..125C}, that is 25\% of all pulsars
in MW GCs. \citet{2006ApJ...651.1098H} detected 50 X-ray sources with Chandra
data. The distance of Terzan\,5 was derived in \citet{2007A&A...470.1043O} by
comparing the horizontal branch (HB) level relative to the reddening lines
over the HB of the template cluster NGC\,6528, using NICMOS and SOFI
near-infrared (NIR) photometry.

RR Lyrae stars (RRLs) are radially pulsating stars, characteristic of
metal-poor, old (Population II) stellar populations.
With periods ranging from 0.2 to 1.0 days, these variable stars are more
commonly found on the instability strip of metal-poor GCs ($\rm{[Fe/H]}
\lesssim -0.8$).
Assuming metallicity and reddening values for Galactic GCs, the cluster
distances can be precisely derived with well-calibrated period--luminosity
and luminosity--metallicity relations \citep[e.g.][]{2017A&A...605A..79G,
2021MNRAS.502.4074M}. The combination of the absolute magnitudes with
the mean magnitudes of the RRLs, observed with time-series photometry,
makes them useful standard candles. 

In this work, we derive accurate distances for the bulge GCs NGC\,6266,
NGC\,6441, NGC\,6626, NGC\,6638, NGC\,6642, and NGC\,6717, for which new data
of the fourth release of the Optical Gravitational Lensing Experiment
(OGLE-IV) survey revealed larger samples of RRLs \citep{2019AcA....69..321S}.
These six moderately metal-poor GCs are within the selection of bulge GCs by
\citet{2016PASA...33...28B}, and they are located in the direction of the
Galactic centre
($R_{\rm GC} < 4.0\,\rm{kpc}$), with a relatively high foreground reddening of
$E(B-V)\sim0.40$. 

In addition to the high stellar crowding present in the bulge, the high total
and differential extinctions hamper the distance derivation through isochrone
fitting even more since they produce a non-uniform spread in
colour-magnitude diagrams \citep[CMDs; e.g.][]{2011AJ....141..146A}. These
effects are mitigated when going to redder wavelengths, such as the \textit{I}
or near-infrared bands, for which $A_I/A_V\sim 0.60$ and $A_{K_S}/A_V \sim
0.12$ \citep[e.g.][]{2019A&A...627A.145O,2019MNRAS.484.5530K}. In particular
for the study of RRLs, the \textit{I} filter gives the best compromise between
spatial resolution and light curve amplitude, which is around $0.3-0.8$\,mag
(compared to $0.5-1.0$\,mag for the \textit{V} filter). For this reason, the
mean \textit{I} magnitudes provided by the new data from OGLE-IV \citep
{2019AcA....69..321S} greatly contribute to complete the census of RRLs in
bulge GCs.

In order to gather the relevant data on the RRLs of the sample clusters, we
cross-matched the recent catalogues of RRLs from OGLE-IV with the earlier
catalogues from \citet[2017 edition]{2001AJ....122.2587C} and \textit{Gaia}
Data Release 2 \citep[DR2;][]{2016A&A...595A...1G,2018A&A...616A...1G}.
We also cross-identified the RRLs with the absolute proper motions (PMs) from
\textit
{Gaia} Early Data Release 3 \citep[EDR3;][]{2021A&A...649A...1G}, and
assigned a membership probability to select a reliable sample of cluster
RRLs. These data allow one to update globular cluster distances with a relatively high
accuracy and better constrain the free parameters contained in an isochrone
fitting \citep[e.g.][]{2019MNRAS.484.5530K,2020ApJ...891...37O,
2020ApJ...890...38S}.
This approach based on member RRLs is not affected as the isochrone fitting
techniques by the problems of binarity, field contamination, and distortion of
the CMD due to the dependence of reddening correction on the stellar effective
temperature, which can dramatically hamper the possibility to obtain reliable
distances of very reddened, low Galactic latitude clusters.

It is worth noting that NGC\,6266, NGC\,6441, and NGC\,6626 host: seven, six
and 14 pulsars\footnote{\url{https://naic.edu/~pfreire/GCpsr.html}},
respectively. In
this sense, a precise age and distance derivation for these GCs have a crucial
importance, since their distances can be used to compute stellar densities and
stellar interaction rates \citep{2007A&A...470.1043O}. 

A recent effort on distance derivation from \textit{Gaia} DR2 PMs and radial
velocities was carried out in \citet{2019MNRAS.482.5138B}, where they
calculated the kinematic distances of 154 GCs by fitting \textit{N}-body models
with a maximum-likelihood approach. Comparing their findings with \citet[2010
edition\footnote{\url{http://physwww.mcmaster.ca/~harris/mwgc.dat}}, hereafter
\citetalias{1996AJ....112.1487H}]{1996AJ....112.1487H} and \citet
{2015ApJ...803...29W}, they found a good agreement for distances up to $\sim
7$\,kpc, but derived systematic 10\% higher distances beyond it. More
recently, \citet{2021MNRAS.505.5978V} and \citet{2021MNRAS.505.5957B} derived
distances from \textit{Gaia} EDR3 parallaxes, velocity dispersion profiles, and
stellar counts, and they compared them to an average of literature distances.
For our sample, the average resulted in 6.41\,kpc, 12.73\,kpc, 5.37\,kpc,
9.78\,kpc, 8.05\,kpc, and 7.52\,kpc, whereas the parallaxes resulted in
5.55\,kpc, 12.66\,kpc, 5.10\,kpc, 9.01\,kpc, 8.26\,kpc, 8.85\,kpc
for NGC\,6266, NGC\,6441, NGC\,6626, NGC\,6638, NGC\,6642, and NGC\,6717,
respectively. 

Several literature works on these GCs used the distances from \citetalias
{1996AJ....112.1487H} as input parameters, using in turn an average of the
$V_{\rm HB}$ magnitudes measured in the literature with isochrone fitting
methods as a distance indicator. In the case of NGC\,6266, \citetalias
{1996AJ....112.1487H} reports the $V_{\rm HB}$ from \citet
{1996A&A...311..778B}, who give $6.8$\,kpc. Following works derived
slightly different distances: $6.95$\,kpc \citep{1999AJ....118.1738F}; $6.64
\pm0.51$\,kpc \citep{2006AJ....131.2551B}; $6.67\pm0.44$\,kpc \citep
{2010AJ....140.1766C}; $7.05\pm0.58$\,kpc \citep{2011AJ....142..163M}; and
$6.42\pm0.14$\,kpc \citep{2015ApJ...803...29W}. For a long time, NGC\,6266 has
been known to host several RRLs \citep{1902AnHar..38....1B,
1959AnLei..21..253V}, and most of them were discovered by \citet
{2010AJ....140.1766C}, placing it as the second cluster with the highest
number of RRLs, only below NGC\,5272.

Despite its high metallicity \citep{1997ApJ...484L..25R}, the dense cluster
NGC\,6441 contains an extended blue HB with a sizable population of peculiar
RRL stars. For this cluster, \citetalias{1996AJ....112.1487H} reported a
distance of $10.4-11.9$\,kpc from \citet{2001AJ....122.2600P}, and the
following works have provided 13.5\,kpc from SOFI@NTT NIR data \citep{2007AJ....133.1287V},
13.61\,kpc from a K-band period--luminosity--metallicity relation \citep
{2008MmSAI..79..355D}, and 13.0\,kpc from VVV NIR data of RRLs
\citep{2021A&A...651A..47A}. Apart from the strong differential reddening in
the bulge region, this discrepancy can be explained by an overabundance of
\element{He} which is usually neglected, as discussed in \citet{2021A&A...651A..47A}.

For NGC\,6626 (M28), \citetalias{1996AJ....112.1487H} reported the HB magnitude
from \citet{2001AJ....121..916T} as $V_{\rm HB}=15.55\pm0.10$ \citep
[close to 15.5 from][]{1996ApJ...468..641D}, a distance of $5.5$\,kpc.
Recently, \citet{2018ApJ...853...15K} derived a distance of $5.34 \pm 
0.21$\,kpc by applying statistical isochrone fitting to \textit{HST}
proper-motion-cleaned CMDs, and \citet{2021A&A...651A..47A} derived
$5.41$\,kpc.
For NGC\,6638, \citetalias{1996AJ....112.1487H}
reported a distance of $9.4$\,kpc from \citet{2002A&A...391..945P}, which was
derived with
{\it HST}/WFPC2 photometry. \citet{2005MNRAS.361..272V} obtained $10.33$\,kpc
from SOFI NIR data, coherent with the results from \citet{2021MNRAS.505.5957B}.

For NGC\,6642, \citetalias{1996AJ....112.1487H} reported 8.1\,kpc from \citet
{2002A&A...391..945P}. Other works derived the following: $7.2\pm0.5$\,kpc using SOAR {\it
BVI} photometry \citep{2006A&A...449.1019B}; 8.63\,kpc from SOFI NIR data
\citep{2007AJ....133.1287V}; and $8.05\pm0.66$\,kpc from \textit{HST}/ACS
photometry \citep{2009MNRAS.396.1596B}. Finally, for NGC\,6717, \citetalias
{1996AJ....112.1487H} reported 7.1\,kpc from \citet{1999A&AS..136..237O},
which was derived with \textit{BV} photometry from the Danish telescope. With {\it
HST}/ACS
data and different isochrones, \citet{2010ApJ...708..698D} and \citet
{2013ApJ...775..134V} obtained 7.55\,kpc and 7.27\,kpc. Very recently, \citet
{2020ApJ...891...37O} derived $7.33 \pm 0.12$\,kpc from \textit{HST} data, by
applying a statistical isochrone fitting method with prior distributions on
the apparent distance moduli, derived from RRL mean magnitudes. 

This paper has the following structure. Section~\ref{sec:LitValues} details
the relevant literature values assumed in the calculations. Section~\ref
{sec:DataCats} presents the different data combined into a single catalogue
of RRLs. Section~\ref{sec:methods} gives the obtained average of the mean
magnitudes, $M_V-$ and $M_I-\rm{[Fe/H]}$ relations and reddening equations. In
Sect.~\ref{sec:results-dist} we discuss the final results on the distances,
compared to recent papers. The conclusions are drawn in Sect.~\ref
{sec:conclusions}. 

\begin{table*}
\caption{Relevant parameters from the literature and input values of
metallicity and reddening adopted in this work.}
\label{tab:Harris}
\centering
\begin{tabular}{lcccccccccc}
\hline\hline
\multirow{2}{*}{Cluster} & $\ell$ & $b$ & $d_{\odot}$ & $R_{\rm GC}$ &
$(m-M)_V$ & Mass & $\rm{[\alpha/Fe]}$& \multicolumn{3}{c}{Inputs} \\
\cline{9-11}
& (deg) & (deg) & (kpc) & (kpc) & (mag) & ($10^5$\,M$_{\odot}$) & (dex) &
$\rm{[Fe/H]}$ & Ref. & $E(B-V)$ \\
\hline
NGC\,6266 & 353.57 & 7.32 & 6.8 & 1.7 & 15.63 & 6.90 & $+0.31$ &
$-1.075\pm0.039$ & L15 & $0.47\pm0.05^\dagger$ \\ 
NGC\,6441 & 353.53 & -5.01 & 11.6 & 3.9 & 16.78 & 12.5 & $+0.28$ &
$-0.50\pm 0.06$ & O08 & $0.47\pm0.05^\dagger$ \\ 
NGC\,6626 & 7.80 & $-5.58$ & 5.5 & 2.7 & 14.95 & 2.84 & $+0.38$ &
$-1.287\pm 0.048$ & V17 & $0.43\pm0.04^\dagger$ \\
NGC\,6638 & 7.90 & $-7.15$ & 9.4 & 2.2 & 16.14 & 1.89 & --- &
$-0.99\pm0.10$ & C09 & $0.42\pm0.04$ \\ 
NGC\,6642 & 9.81 & $-6.44$ & 8.1 & 1.7 & 15.79 & 0.645 & --- &
$-1.19\pm0.10$ & C09  & $0.40\pm0.08$ \\ 
NGC\,6717 & 12.88 & $-10.90$ & 7.1 & 2.4 & 14.94 & 0.181 & --- &
$-1.26\pm 0.10$ & C09 & $0.23\pm0.02$ \\ 
\hline
\end{tabular}
\tablefoot{Coordinates, distances and distance modulus are extracted
from \citet[2010 edition]{1996AJ....112.1487H}, and masses are from \citet
{2018MNRAS.478.1520B}. References for the metallicity: L15 - \citet
{2015ApJ...813...97L}; V17 - \citet{2017MNRAS.464.2730V};
O08 - \citet{2008MNRAS.388.1419O}; C09 - \citet{2009A&A...508..695C}.
References for the reddening are given in the text. \\ $^\dagger$ A small reddening correction, based on the observed and absolute
\textit{VI} magnitudes, will be applied (see Section~\ref{sec:reddening}).}
\end{table*}


%

\section{Metallicity and reddening from the literature}
\label{sec:LitValues}

In order to properly estimate the cluster distances, we need to assume
metallicity and reddening values from the literature, giving preference to
those with the smaller, more reliable uncertainties. The metallicity is
required to be applied in the $M_V-$ and $M_I-\rm {[Fe/H]}$
relations, providing the absolute magnitude $M_V$ and $M_I$, whereas the
foreground reddening $E(B-V)$ is used to convert the apparent distance
moduli $(m-M)_V$ and $(m-M)_I$ into an absolute scale, that is $(m-M)_0$.

Table~\ref{tab:Harris} gives the relevant parameters of the six sample GCs,
including the Galactic coordinates, distances, and distance moduli from
\citetalias{1996AJ....112.1487H}, and masses from \citet
[updated version\footnote{
\url{https://people.smp.uq.edu.au/HolgerBaumgardt/globular/parameter.html}}]
{2018MNRAS.478.1520B}.
The last columns also give the adopted values and uncertainties
of metallicity (with references) and reddening (except
for a small correction that is to be applied in the end of Sect.~\ref{sec:reddening}
for three GCs), which is elucidated below.

For the three clusters with a metallicity available from high-resolution
spectroscopy of individual stars, we
adopted these values with a $3\sigma$ uncertainty: \citet{2015ApJ...813...97L}
for NGC\,6266, \citet{2008MNRAS.388.1419O} for NGC\,6441, and \citet
{2017MNRAS.464.2730V} for NGC\,6626. These three clusters present an
$\alpha$-enhancement of $[\alpha/\rm{Fe}]\sim0.30-0.40$, which is typical for
the
old GC population present in the Galactic bulge. For the remaining clusters,
we adopted the metallicity from \citet{2009A&A...508..695C}, with an average
uncertainty of $0.10\,\rm{dex}$.

Regarding the reddening values, those compiled in \citetalias
{1996AJ....112.1487H} are an average of the values from \citet
{1985IAUS..113..541W}, \citet{1985ApJ...293..424Z}, and \citet
{1988PASP..100..545R}, plus the references given in Sect.~\ref{sec:intro}
\citep{1996A&A...311..778B,2001AJ....122.2600P,2001AJ....121..916T,
2002A&A...391..945P,1999A&AS..136..237O} for each cluster. According to
\citetalias{1996AJ....112.1487H}, the typical errors are around 10\%, but not
lower than $0.01$\,mag. Assuming a unique value of colour excess
may produce a higher dispersion in the distance moduli of the stars in GCs
with significant spatial differential reddening \citep[e.g. NGC\,6266 with
$\Delta E(B-V)\sim0.25$;][]{2012AJ....143...70A}.

For NGC\,6266 and NGC\,6441, we adopted the reddening from \citetalias
{1996AJ....112.1487H}, $E(B-V) = 0.47$, since all the other works are based on
it. For NGC\,6626, we adopted $0.43\pm0.04$, which is an average value given
in \citet
{2018ApJ...853...15K}, derived from two different isochrone models. For
NGC\,6638, we applied $0.42\pm0.04$, which is an average value between \citet{2005MNRAS.361..272V} and \citetalias{1996AJ....112.1487H}. For NGC\,6642, we adopted $0.40\pm0.08$
with a higher uncertainty, given the discrepancy in the literature:
0.40 \citepalias{1996AJ....112.1487H}, 0.42 \citep{2006A&A...449.1019B}, 0.60
\citep{2007AJ....133.1287V}, and 0.43 \citep{2009MNRAS.396.1596B}. For
NGC\,6717, we adopted $0.23$ from \citet{1999A&AS..136..237O} with a 10\%
uncertainty, very close to the average value from \citetalias
{1996AJ....112.1487H}. \citet{2020ApJ...891...37O} found a slightly lower
reddening of $0.19\pm0.02$.


%

\section{RR Lyrae data: OGLE-IV, Clement, and \textit{Gaia}}
\label{sec:DataCats}

Based on the OGLE-IV survey, \citet{2014AcA....64..177S} released the OGLE
Collection of Variable Stars\footnote
{\url{http://ogledb.astrouw.edu.pl/~ogle/OCVS/}} (OCVS), with $\sim38\,000$
RRLs spanning over $182\,\rm{deg}^2$ towards the Galactic bulge. However, this
footprint is restricted to a very central region ($|\ell|\lesssim7^{\circ}$, 
$|b|\lesssim6^{\circ}$), containing around half of the bulge GCs identified so
far. The new observations, described in \citet{2019AcA....69..321S},
extended the covered area to about $3\,000\,\rm{deg}^2$, including a much
larger part of the bulge ($|\ell|\lesssim20^{\circ}$, $|b|\lesssim15^{\circ}$)
and a good extension into the MW disk. According to \citet
{2019AcA....69..321S}, the OCVS now includes 27 GCs hosting RR Lyrae. 

These new observations are part of the OGLE project named the Galaxy
Variability Survey and are shallower than the original OGLE data, with exposure
times of 25\,s instead of $100-150$\,s. \citet{2019AcA....69..321S} report
all the identified RRLs as new, but in the sense that they are new in the OCVS
database, where the number of RRLs more than doubled compared to the previous
data. A further cross-matching with all the other catalogues of RRLs in the
literature is carried out here to check whether the detected RRLs are new.

\begin{figure}
   \centering
   \includegraphics[width=\columnwidth]{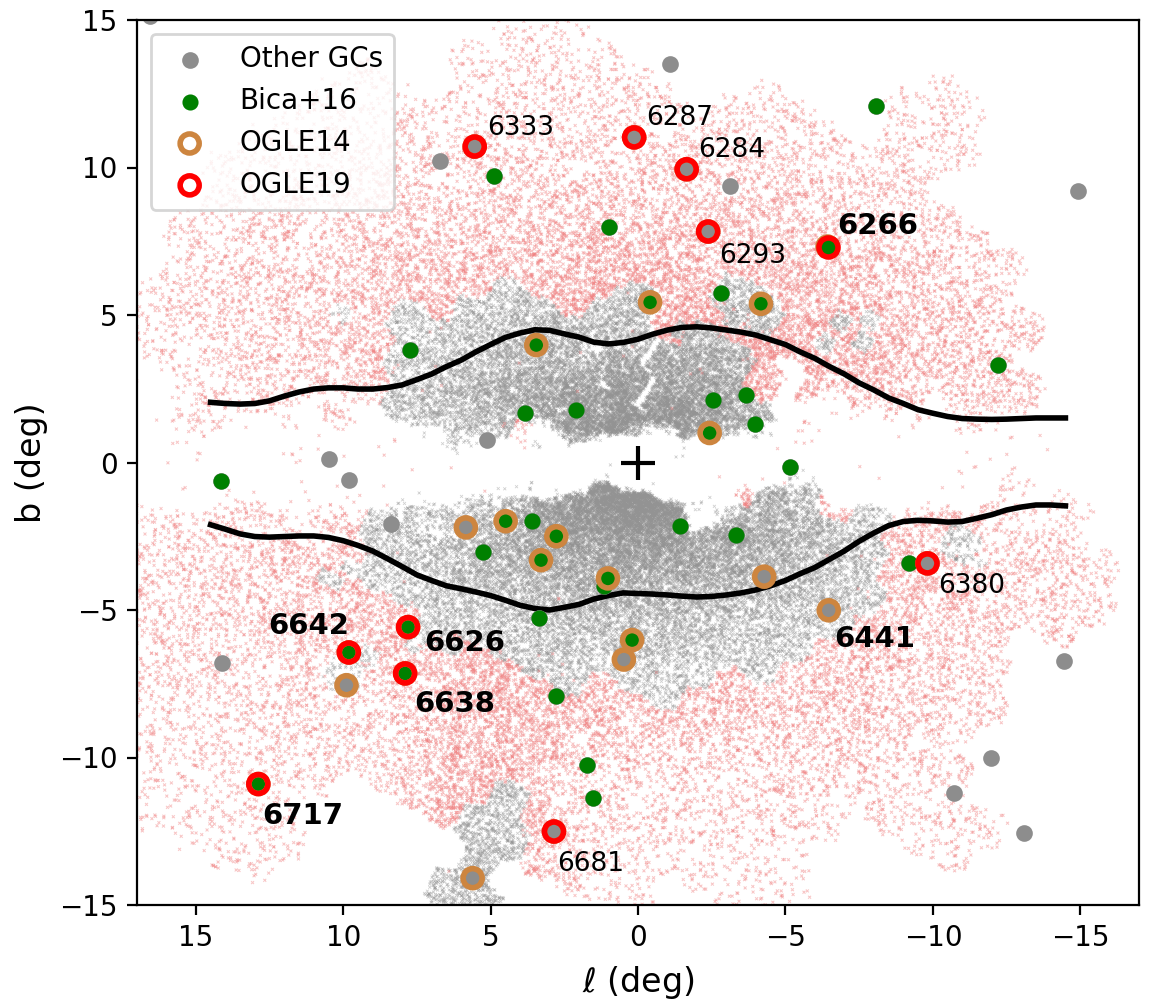}
      \caption{Galactic coordinates of the bulge region, with the black
      contours as the COBE/DIRBE outline \citep{2017A&A...598A.101J}. The
      small symbols are all RRLs from the OGLE-IV survey (in grey from \citealp
      {2014AcA....64..177S}, and in red from \citealp{2019AcA....69..321S}).
      The GCs are also shown, with those selected by \citet
      {2016PASA...33...28B} as bulge GCs shown in green. The GCs with contours
      are the
      ones that contain member RRLs according to \citet{2014AcA....64..177S,
      2019AcA....69..321S}. The selected sample of six clusters consists of
      the objects in common in \citet{2016PASA...33...28B} and \citet{2019AcA....69..321S}, except for NGC\,6441.}
   \label{Fig:lb-RRL}
\end{figure}

\begin{figure*}
   \centering
   \includegraphics[width=0.995\textwidth]{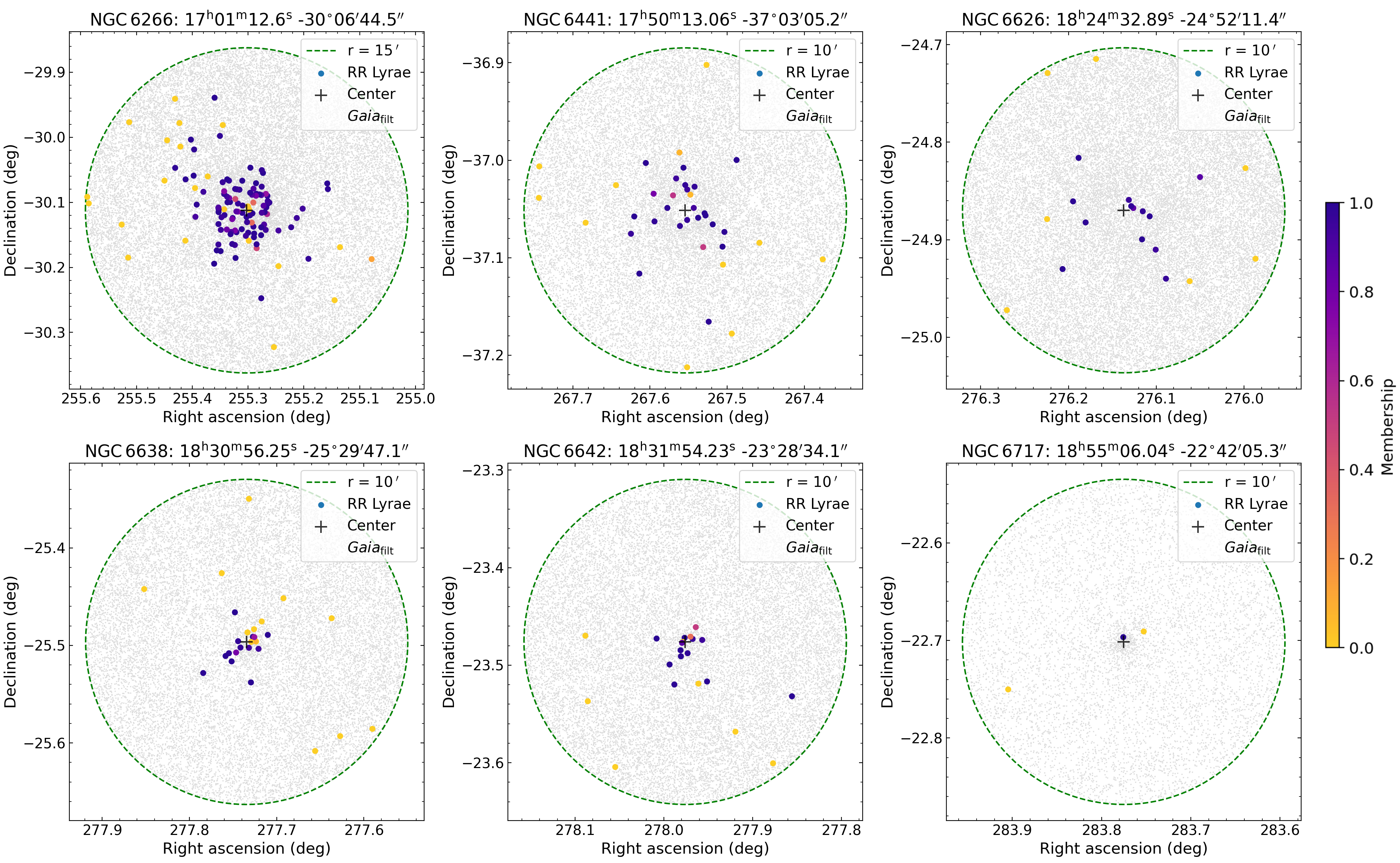}
      \caption{Equatorial coordinates of the RRLs ($N_{\rm filt}$) located inside a radius
      of $15^{\prime}$ around NGC\,6266, and $10^{\prime}$ for the other
      five GCs. The computed membership is represented
      by the colour bar, between 0 and 1. The stars from \textit{Gaia} EDR3,
      filtered by the catalogue validation \citep{2021A&A...649A...8G, 2021A&A...649A...5F, 2021A&A...649A...3R} are shown
      as grey symbols in the background.}
  \label{Fig:RADECmemb}
\end{figure*}

Figure~\ref{Fig:lb-RRL} shows the Galactic coordinates of the $38\,257$
RRLs from \citet{2014AcA....64..177S}, and the $\sim 29\,000$ bulge RRLs from
\citet{2019AcA....69..321S}. All the Galactic GCs from \citetalias
{1996AJ....112.1487H} are overplotted, with those reported in \citet
{2016PASA...33...28B} as bulge GCs shown in green. We also marked the clusters
that contain member RRLs in \citet{2014AcA....64..177S,2019AcA....69..321S}.
The present sample of six GCs corresponds to the unique bulge GCs with member
RRLs in \citet{2019AcA....69..321S}, except for NGC\,6441 located in
an outer bulge shell \citep{2016PASA...33...28B}.
A larger sample of clusters will be analised in a future work.

\begin{table}
\caption{Number of RRLs from \citet{2001AJ....122.2587C} and OGLE-IV
catalogues, and retrieved in each stage of the methods referred to in
Sect.~\ref{sec:methods}.}
\label{tab:ogleclement}
\centering
\begin{tabular}{l c c c c c c}
\hline\hline
Cluster & $N_{\rm Clem}$ & $N_{\rm OGLE}$ & $N_{\rm new}$ & $\mathbf{N_
{tot}}$ & $N_{Gaia}$ & $N_{\rm filt}$\\
\hline
NGC\,6266 & 230 & 231 & 12 & \textbf{242} & 233 & 123 \\
NGC\,6441 &  81 &  56 & 11 & \textbf{ 92} &  87 &  37 \\
NGC\,6626 &  22 &  27 & 11 & \textbf{ 33} &  32 &  20 \\
NGC\,6638 &  28 &  29 & 13 & \textbf{ 41} &  41 &  28 \\
NGC\,6642 &  17 &  29 & 13 & \textbf{ 30} &  29 &  21 \\
NGC\,6717 &   1 &   4 &  3 & \textbf{  4} &   4 &   3 \\
\hline
\end{tabular}
\tablefoot{Number of RRLs from \citet
[$N_{\rm Clem}$]{2001AJ....122.2587C} and OGLE-IV ($N_{\rm OGLE}$); number of
new RRLs, present in OGLE but not in Clement ($N_{\rm new}$); number of RRLs
in the combined catalogues ($N_{\rm tot}$); number of RRLs detected in
\textit{Gaia} EDR3 \citep{2021A&A...649A...1G} before and after applying the
quality flags ($N_{Gaia}$ and $N_{\rm filt}$).}
\end{table} 

From the updated OCVS database, we retrieved a list of RRLs with the \textit{I}
mean magnitude, located inside a circular area around the cluster centre:
$15^{\prime}$ for NGC\,6266 \citep[to include all the RRLs from][]
{2001AJ....122.2587C} and $10^{\prime}$ for the other clusters.
The catalogues
were cross-matched in position with the catalogues of \citet[2017
edition\footnote{\url{http://www.astro.utoronto.ca/~cclement/read.html}}]
{2001AJ....122.2587C}, which contain the mean \textit{V} magnitudes.
Table~\ref{tab:ogleclement} gives the number of RRLs retrieved for each cluster
in these two catalogues, the number of OGLE-IV RRLs that are actually new, and
the number of RRLs in the combined catalogue.

\subsection{Gaia EDR3: Positions and proper motions}
\label{subsec:gaia3}

In order to carry out a membership analysis, we accessed the high-precision
astrometric and photometric data from the \textit{Gaia} EDR3. For our samples,
the main improvements compared to the previous DR2 were
the number of detected sources and smaller PM errors. Following the
recommendations of the \textit{Gaia}
collaboration\footnote{\url{https://www.cosmos.esa.int/web/gaia/edr3-code}},
the following four corrections were applied to the data: parallax zero-point
correction
\citep{2021A&A...649A...4L}, G-band magnitude and flux correction, flux
excess factor correction \citep{2021A&A...649A...3R}, and a recent
correction of the PM bias with $G$ magnitude \citep
{2021A&A...649A.124C}. 

\begin{figure*}
   \centering
   \includegraphics[width=\textwidth]{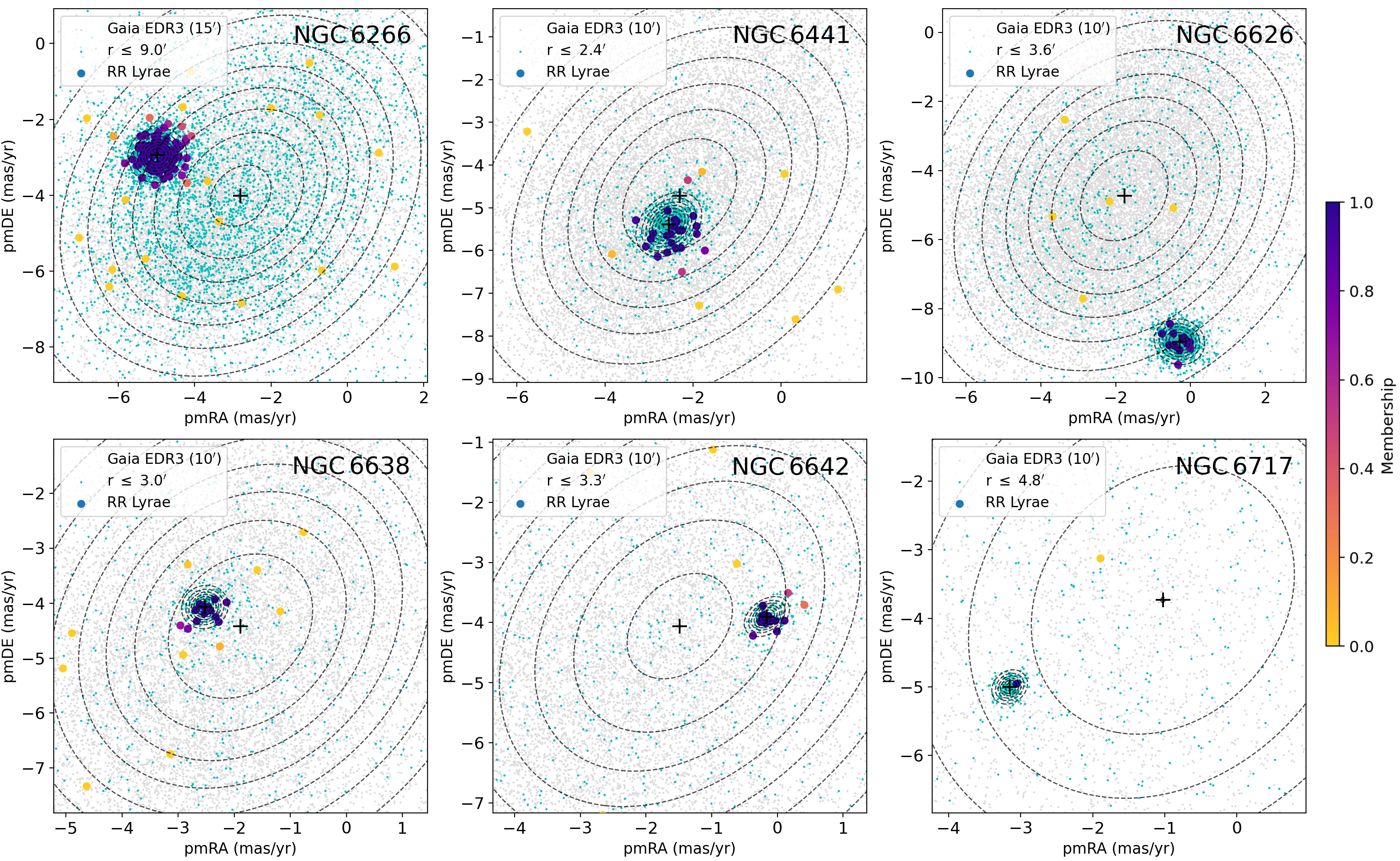}
      \caption{PM diagram of the sample clusters, with the RR
      Lyrae colour-coded by the membership values. The background stars are
      divided into those that are more central and those that are further out,
      shown as cyan and grey points. The two-dimensional Gaussian distributions mark the
      central cluster and field PMs, along with the dispersions
      (contour lines mark $0.25-2\sigma$ in steps of $0.25\sigma$), as
      derived from the GMM
      method. The diagrams in the lower panels are zoomed in due to the
      lower dispersion of the cluster PMs.}
    \label{Fig:pmMemb}
\end{figure*}

As concerns the catalogue validation, four suggested criteria were tested:
$G\leq 19$\,mag \citep{2021A&A...649A...5F}, $G_{\rm RP} \leq 20$\,mag,
|\texttt{phot\_bp\_rp\_excess\_factor}$|<5\sigma_{C^*}$ \citep
{2021A&A...649A...3R}, and re-normalised unit weight error (\texttt{ruwe})
$< 1.4$ \citep{2021A&A...649A...8G}. \citet{2021A&A...649A...3R} show that
the filtering in $\sigma_{C^*}$ removes sources with inconsistencies between
the $G$, BP, and RP photometry, affecting the completeness of variable and
extended sources. In fact, some central RRLs were filtered in these clusters,
and we opted not to use this filtering and to loosen the last
restriction to $\texttt{ruwe}< 2.0$ \citep[since \texttt{ruwe} is higher in
crowded areas;][]{2021A&A...649A...4L}. With these changes, we obtained
filtered samples ($N_{\rm filt}$; Table~\ref{tab:ogleclement}) more similar
to \textit{Gaia} DR2. 



After transforming the \textit{Gaia} ICRS coordinates to J2000, we
cross-matched the \textit{Gaia} EDR3 catalogues, appending the PMs and
\textit{Gaia} magnitudes to our combined catalogue. We applied
a two-dimensional Gaussian mixture model (GMM) to a group of more central
stars against the remaining field stars. The radius limiting this group of
central stars was obtained iteratively as the minimum radius that returns a
convergence of both cluster and field PMs in right ascension and declination
directions. The GMM method assumes the data are clustered in the
parameter space following a superposition of Gaussian distributions and it uses
the expectation-maximisation algorithm to determine the parameters of each
distribution and a correlation matrix \citep{10.5555/1403886}. In this case,
there are two Gaussian distributions (the cluster distribution with a lower
dispersion versus the field stars) in a
two-dimensional PM plane. It was applied using using the \texttt{Python}
library \texttt{scikit-learn} \citep{scikit-learn}. 

A membership probability of the RRLs was computed using the equations from
\citet{2009A&A...493..959B}, which consider the measured PMs of each RRL, the
cluster and the field, and also their respective uncertainties. Figure~\ref
{Fig:RADECmemb} shows the coordinates of the RRLs in the field around the six
sample GCs, colour-coded
with the final membership values. Since the membership was considered
when computing the weighted average of the mean magnitudes (Section~\ref
{subsec:average}), no lower cut was applied and all the RRLs were
maintained in the catalogues. 

The PM diagrams of the six sample clusters are presented in Fig.~\ref
{Fig:pmMemb}, showing the centre and dispersion of the two fitted Gaussian
distributions, and the groups of central and field stars. The computed PMs in
right ascension and declination ($\mu_{\alpha}\cos\delta$ and $\mu_{\delta}$)
are given
in the first columns of Table~\ref{tab:avgMags}. The values are very coherent
with those from \citet{2021MNRAS.505.5978V}, which were also derived with
\textit{Gaia} EDR3 data.

The final RR Lyrae catalogues of the six sample GCs, combining the data from
OGLE-IV, \citet{2001AJ....122.2587C}, and \textit{Gaia} EDR3, along with
the membership probability values, are provided by us via the \texttt{VizieR}
platform. The database can contribute to a wide range of studies by
providing a sample of member RRLs for these clusters and removing foreground
or background ones.

\section{Methods: Distances from OGLE mean magnitudes}
\label{sec:methods}

\begin{table*}
\caption{Derived cluster PMs projected in the right ascension and
declination components (Sect.~\ref{subsec:gaia3}). The weighted average of the
mean \textit{V} and \textit{I} magnitudes are also given with the number of
stars in the calculation before the sigma clipping removed outliers ($N_V$
and $N_I$).}
\label{tab:avgMags}
\centering
\begin{tabular}{l c c c c c c}
\hline\hline
\multirow{2}{*}{Cluster} & $\mu_\alpha\cos\delta$ & $\mu_\delta$ & \multirow{2}{*}{$N_V$}
& $\langle V\rangle$ & \multirow{2}{*}{$N_I$} & $\langle I\rangle$ \\
& (mas\,yr$^{-1}$) & (mas\,yr$^{-1}$) & & (mag) & & (mag) \\
\hline
NGC\,6266 & $-4.978\pm0.062$ & $-2.944\pm0.058$ & 100 & $16.298\pm0.023$ &
123 & $15.169\pm0.014$ \\
NGC\,6441 & $-2.551\pm0.049$ & $-5.393\pm0.051$ &  30 & $17.473\pm0.035$ &
 37 & $16.473\pm0.029$ \\
NGC\,6626 & $-0.296\pm0.051$ & $-8.954\pm0.046$ &  10 & $15.807\pm0.054$ &
 20 & $14.628\pm0.033$ \\
NGC\,6638 & $-2.523\pm0.042$ & $-4.069\pm0.039$ & --- & ---              &
 28 & $15.879\pm0.035$ \\
NGC\,6642 & $-0.161\pm0.042$ & $-3.901\pm0.038$ & --- & ---              &
 21 & $15.431\pm0.032$ \\
NGC\,6717 & $-3.153\pm0.031$ & $-5.001\pm0.030$ &   1 & $15.700\pm0.100$ &
  3 & $14.881\pm0.080$ \\
\hline
\end{tabular}
\end{table*}

In this section, we describe the methods applied to calculate the heliocentric
distances: going from computing the average of the mean magnitudes ($\langle V
\rangle$ from \citealt{2001AJ....122.2587C} catalogues, and $\langle I \rangle$
from \citealt{2019AcA....69..321S}), to the determination of the adequate
$M_V-M_I-\rm{[Fe/H]}$ relations from \textit{a Bag of Stellar
Tracks and Isochrones}\footnote{Available at: \url{http://basti-iac.oa-abruzzo.inaf.it}.}
\citep[BaSTI;][]{2021ApJ...908..102P} models, and a 
discussion on the proper reddening laws to be applied in such a relatively
high-reddening regime.

Even more so than the new RRLs detected in \citet[see Table~\ref
{tab:ogleclement}]{2019AcA....69..321S}, the main contribution of the new
OGLE-IV catalogues to our analysis are the well-calibrated mean $I$
magnitudes, since this analysis is normally done in the $V$ band. However,
from this point on, we proceed with the analysis
of both $V$ and $I$ bands, allowing one to compare the results and
argue about systematic differences.

\subsection{Weighted average of the mean magnitudes}
\label{subsec:average}

The light curves, \textit{I}-band amplitudes, and mean magnitudes given in the
new OGLE-IV catalogues are a combination of $20-200$ exposures of $25$\,s, with
a median value of 112 epochs \citep{2019AcA....69..321S}. The data are part of
the Galaxy Variability Survey, which are $\sim1$\,mag shallower than the
original OGLE photometry. According to \citet{2019AcA....69..321S}, the
photometric saturation limit is $I\sim11$\,mag and the faint limit is
$I\sim19.5$\,mag. This magnitude range is enough to study the RRLs and HB
of several bulge GCs since they populate the CMDs around $I\sim15$
(or $V \sim 16$) in this reddening regime.

We calculated an average of these mean magnitudes, which are quite stable
around the RRL locus of old GCs. As mentioned before, the averages $\langle V
\rangle$ and $\langle I \rangle$ are based on the mean $V$ and $I$ magnitudes
by \citet{2001AJ....122.2587C} and \citet{2019AcA....69..321S}. These averages
were calculated  with a weight for each RRL, corresponding to its
memberships $p_i$ and $p_j$: 
\begin{equation}
    \centering
    \ \langle V\rangle = \frac{\sum_i\, \langle V \rangle_i p_i}{\sum_i\, p_i}
    \;\;\;\mathrm{and}\;\;\; \langle I\rangle = \frac{\sum_j\, \langle I\rangle_j
    p_j}{\sum_j\, p_j}\;\;,
\end{equation}
where the indexes $i$ and $j$ of the two summations go from 1 to $N_V$ and
$N_I$, respectively.

Moreover, we applied a sigma clipping to iteratively remove the outliers
located beyond $2\sigma$ from the median magnitude, until a convergence was
reached, where $\sigma$ is the standard deviation. Since the outliers were
mostly low-$p_i$ RRLs, the central results do not change much, but the final
standard error of the weighted mean is reduced by half due to the smaller
dispersion of the adopted points. The sigma clipping is crucial because the
magnitude of the outliers could be scattered by physical double stars or by
other peculiar errors.

\begin{figure}
    \centering
    \includegraphics[width=\columnwidth]{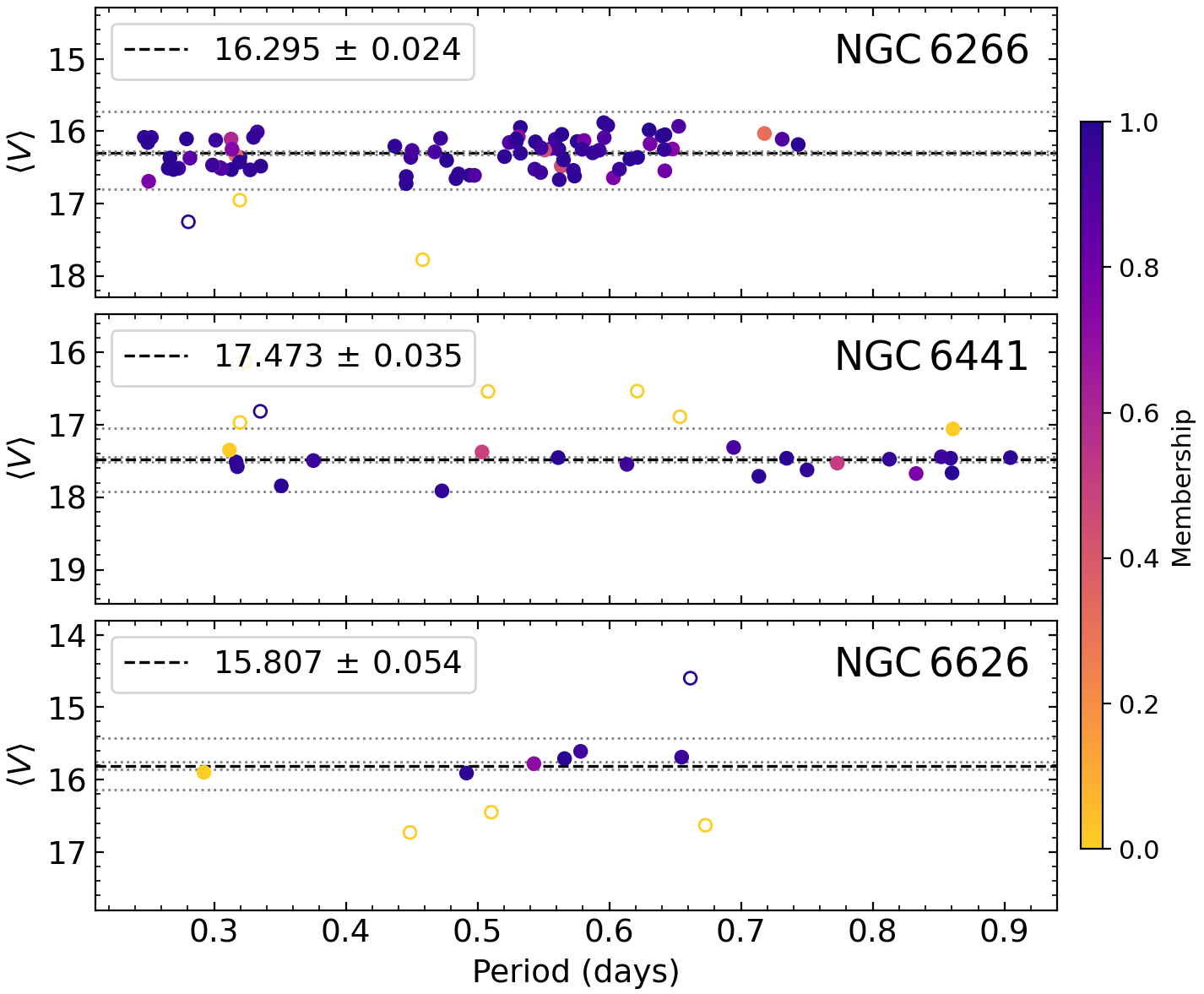}
    \caption{Mean \textit{V} magnitude \citep{2001AJ....122.2587C} versus
    period of pulsation for the RRLs of NGC\,6266, NGC\,6441, and NGC\,6626.
    The stars are colour-coded by the derived membership, and the empty
    symbols correspond to stars that were removed using the sigma clipping
    method ($\rm{median} \pm 2\sigma$). The dashed black line represents the
    weighted average, and the dotted lines give the standard error and the
    $2\sigma$ level.}
    \label{fig:AvgMags}
\end{figure}

\begin{figure}
    \centering
    \includegraphics[width=\columnwidth]{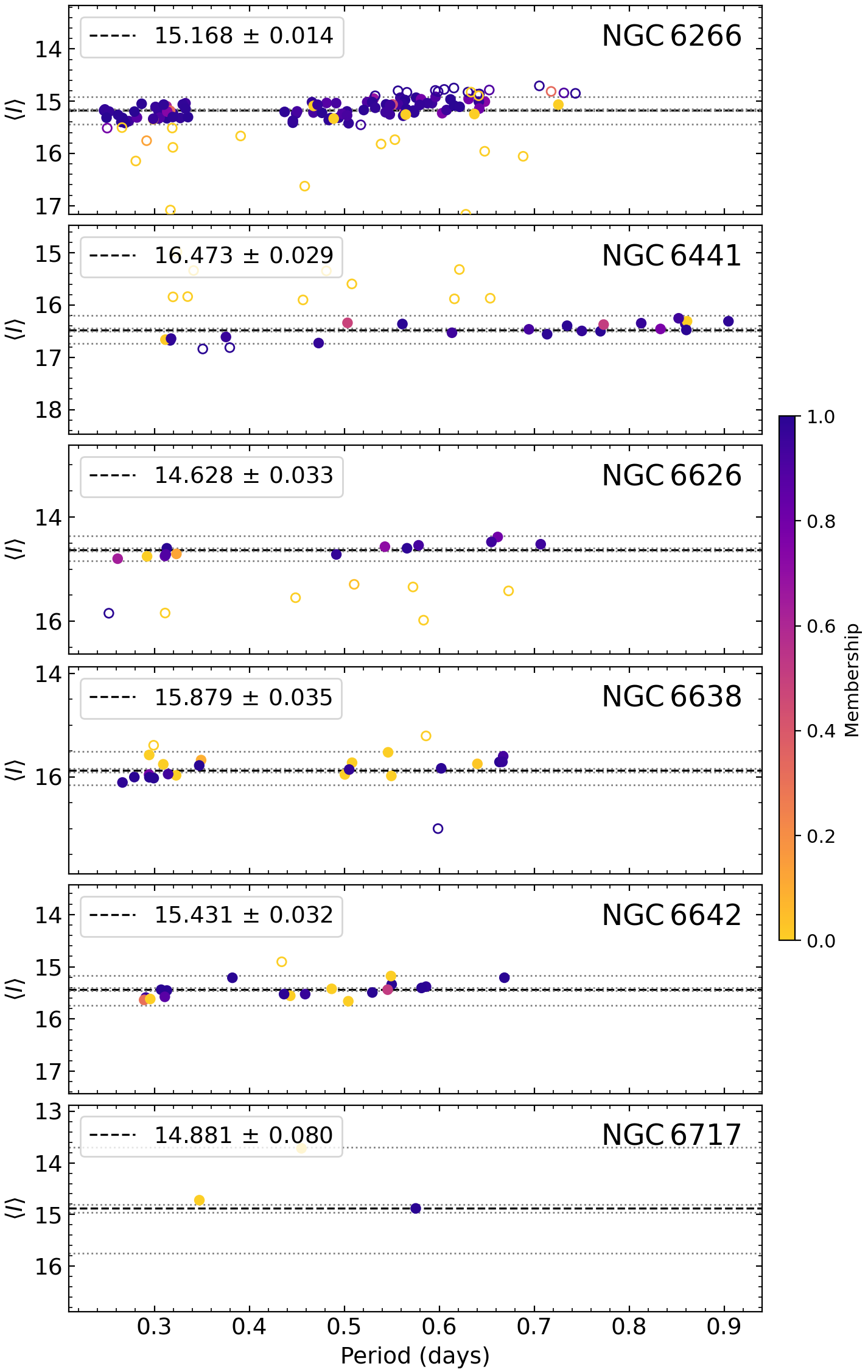}
    \caption{Mean \textit{I} magnitude \citep{2019AcA....69..321S} versus
    period for the six sample clusters. The details are the same as in
    Fig.~\ref{fig:AvgMags}.}
    \label{fig:AvgMags2}
\end{figure}

The results on the weighted average of mean $V$ and $I$ magnitudes, along with
the standard error and the number of stars considered in each case, are also
given in Table~\ref{tab:avgMags}. The number of stars ($N_V$ and $N_I$)
corresponds to the stars with available mean magnitudes which are present in
the filtered \textit{Gaia} EDR3 catalogues,
excluding those
with PM errors greater than $0.20$\,mas\,yr$^{-1}$. Figures~\ref{fig:AvgMags}
and \ref{fig:AvgMags2} show the $\langle V\rangle$ and $\langle I \rangle$
versus period plots for the clusters with more than one member RRL with
available magnitudes, with the RRLs colour-coded by its membership results.

It is clear from Fig.~\ref{Fig:RADECmemb} that the radius alone is not a valid
parameter for selecting member RRLs since several central RRLs have a low
PM-based membership, and vice versa. Figures~\ref{fig:AvgMags} and \ref
{fig:AvgMags2} also show that some RRLs with a low membership have apparent
magnitudes close to the weighted average, that is they have the same distance
as the cluster. Therefore, it is possible that some real RRL members returned
low membership values, but it is also expected that the large number of RRLs
reduces this bias.


\subsection{$M_V-$ and $M_I-$[Fe/H] relations from BaSTI $\alpha$-enhanced
models}
\label{subsec:basti}

Several luminosity-metallicity relations are available in the literature with
slightly different slopes \citep[e.g.][]{1993AJ....106..703S,
2003AJ....125.1309C, 2017A&A...605A..79G}. The relations between absolute
magnitude and metallicity are normally given in the $V$ band ($M_V-\rm
{[Fe/H]}$), whereas in the near- and mid-infrared a period--luminosity or
period--luminosity--metallicity relation is obtained \citep
[e.g. $PM_{K_S}Z$ in][]{2018MNRAS.481.1195M}. Most of them employ
nearby field RRLs, possibly Population I stars, younger than our very
old, metal-poor, $\alpha$-enhanced bulge sample.

Large samples of RRLs with accurate trigonometric parallaxes are required to
calibrate (period--)luminosity--metallicity relations. Such a large sample
was made possible with the recent \textit{Gaia} mission, containing around
400 MW RRLs. Here, we compare the following two recent works that derived
$M_V-\rm{[Fe/H]}$, along with other relations, adopting the \textit{Gaia}
parallaxes: \citet{2017A&A...605A..79G} and \citet{2018MNRAS.481.1195M}.

\citet{2017A&A...605A..79G} calibrated $M_V-\rm{[Fe/H]}$ relations based on
parallaxes from the Tycho-\textit{Gaia} Astrometric Solution (TGAS) from
\textit{Gaia} DR1 \citep{2016A&A...595A...2G}. They applied three
different methods to perform this calibration, fixing the slope in $0.214\pm
0.047$\,mag/dex, as determined in two earlier photometric and spectroscopic
studies of RRLs in the Large Magellanic Cloud bar \citep{2003AJ....125.1309C,
2004A&A...421..937G}, returning $M_V = 0.214\rm{[Fe/H]} + 0.88^{+0.04}
_{-0.06}$. 

On the other hand, \citet{2018MNRAS.481.1195M} derived $M_V-\rm{[Fe/H]}$ by
adopting photometric data from \citet{2013MNRAS.435.3206D} and parallaxes from
\textit{Gaia} DR2, obtaining $M_V=(0.34\pm0.03)\cdot\rm{[Fe/H]}+(1.17\pm0.04)$
for 381 RRLs. This higher slope is closer to those from \citet
{1993AJ....106..703S} and \citet{1997MNRAS.284..761F}, but steeper than that
from \citet{2017A&A...605A..79G}. This difference can be due to the zero-point
offset affecting the \textit{Gaia} DR2 parallaxes \citep{2018A&A...616A..17A}
and the assumed metallicity values from \citet{2013MNRAS.435.3206D}. When
limiting their study to a smaller sample of 23 RRLs with a metallicity from
high-resolution spectroscopy, \citet{2018MNRAS.481.1195M} obtained $M_V =
(0.25\pm0.05) \rm{[Fe/H]} + (1.18 \pm0.12)$, which is much closer to the
recent literature.

For the present work, we decided to estimate the distance to the
selected clusters by using the theoretical distance scale based on the most
updated set of BaSTI stellar models \citep{2018ApJ...856..125H,
2021ApJ...908..102P}. In more detail, since we were faced with metal-poor,
$\alpha$-enhanced star clusters, we selected the $\alpha$-enhanced version of
the BaSTI library recently provided by \citet{2021ApJ...908..102P}: the
predicted magnitudes in the \textit{VI} Johnson-Cousin bands of zero-age
horizontal branch (ZAHB) models located within the RRL instability strip
($\log T_{\rm eff}=3.83$) were used to derive the dependence of the HB
brightness (in these specific photometric passbands) on the metallicity.
Since we are interested in comparing the average magnitude of the RRL sample
in each cluster with the theoretical predictions, we corrected the ZAHB
brightness by applying a calibration from \citet{1997MNRAS.285..593C} in order
to properly account for the post-ZAHB evolutionary effects that impact on
the RRL luminosity distribution \citep[we refer to][for a detailed discussion
on this issue]{2004A&A...426..641C}.





\begin{figure}
    \centering
    \includegraphics[width=0.935\columnwidth]
    {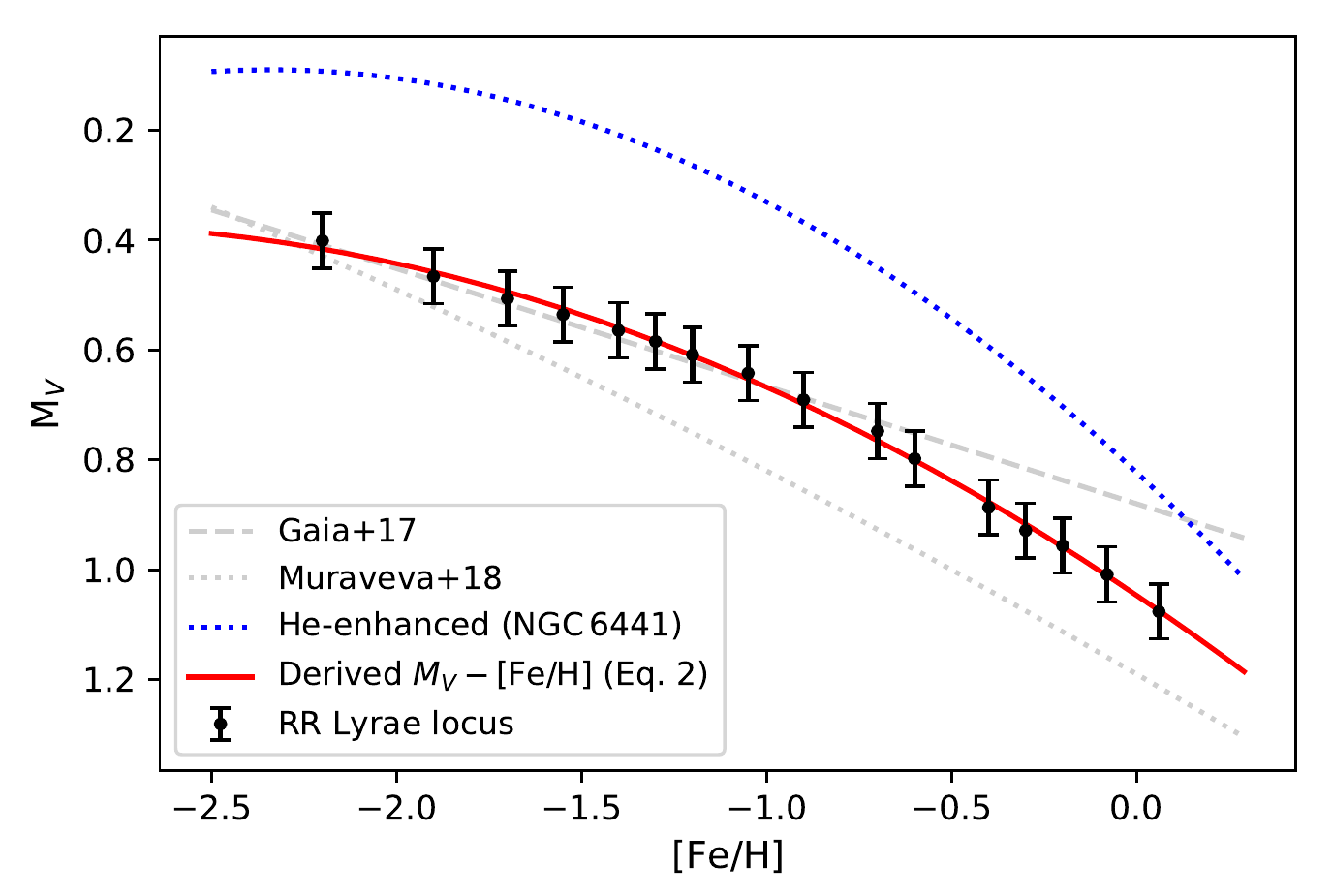}
    \includegraphics[width=0.935\columnwidth]
    {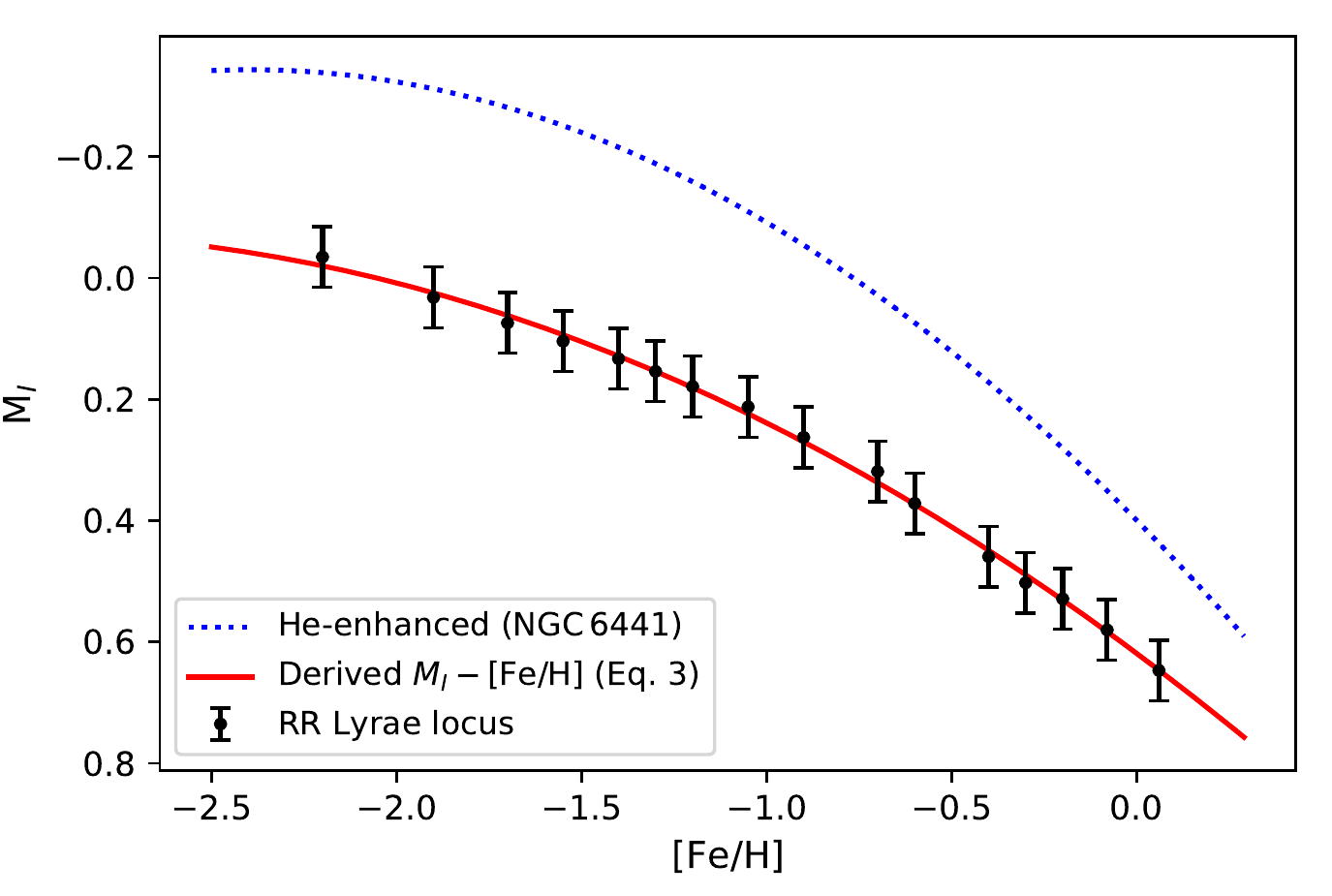}
    \caption{Quadratic fits of the luminosity-metallicity relations
    $M_V-\rm{[Fe/H]}$ and $M_I-\rm{[Fe/H]}$. The black points are BaSTI
    $\alpha$-enhanced
    zero-age HB models, corrected by the calibration from \citet
    {1997MNRAS.285..593C} to account a dispersion of the RRLs due to their
    evolution. The quadratic functions (red for canonical He, as in
    Eqs.~\eqref{eq:1} and \eqref{eq:2}, and dotted blue
    for enhanced He) were obtained via the Markov chain Monte
    Carlo method.}
    \label{fig:mFEH-plot}
\end{figure}

The GC NGC\,6441 as its twin NGC\,6388, at odds with its metal content,
shows an extended blue HB \citep{1997ApJ...484L..25R}, and hosts a peculiar
class of RRL characterised by anomalously large pulsation periods, defining
the peculiar Oosterhoff class named OoIII \citep{2003AJ....126.1381P}. Detailed
HB simulations \citep{2007A&A...474..105B,2017MNRAS.465.1046T} have shown that
the peculiar HB morphology and the pulsational properties of its RRL population
can be explained by accounting for the presence of a helium-enhanced stellar
population with $Y \sim 0.35-0.40$. With the RRLs in NGC\,6441 being the
progeny of this He-enhanced sub-population, they cross the instability strip
at a larger luminosity, and hence a longer period, than normal He RRLs. To
take this occurrence into account, we adopted He-enhanced BaSTI ZAHB models
with $Y=0.35$ to estimate the RRL strip luminosity level, which is to be
applied for estimating the distance to NGC\,6441.

Figure~\ref{fig:mFEH-plot} shows the absolute magnitudes $M_V$ and $M_I$ versus
the metallicity of the ZAHB models (black points). The metallicity ranges
between
$-2.20$ and $+0.06$, and the vertical error bars are assumed to be of $0.05$.
The coefficients of the quadratic function (red curves), better following
the points than a linear one,
were obtained with a Markov chain Monte Carlo method
\citep[using the \texttt{Python} library \texttt{emcee};][]
{2013PASP..125..306F}, along with the
respective uncertainties. The dotted line shows the relationship
derived for the He-enhanced ZAHB models ($Y=0.35$), adopted for the case of
NGC\,6441.

%
\begin{table*}
\caption{Results of the absolute magnitudes, distance moduli, and
distances obtained from the mean $I$ and $V$ magnitudes and BaSTI ZAHB models.
The distances derived by \citet[BV21]{2021MNRAS.505.5957B} and \citet[VB21]
{2021MNRAS.505.5978V} are also provided for comparison purposes.}
\label{tab:Results-I}
\centering
\begin{tabular}{l c c c c c c c c}
\hline\hline
Cluster & $M_I$ &
$(m-M)_I$ & \textbf{D}$_\mathbf{I}$ [kpc] &
$M_V$ & $(m-M)_V$ & \textbf{D}$_\mathbf{V}$ [kpc] &
D$_{\rm BV21}$ & D$_{\rm VB21}$ \\
\hline
NGC\,6266 & $0.22\pm0.08$ & 
$14.95\pm0.08$ & $\mathbf{6.6\pm 0.4}$ & 
$0.65 \pm 0.08$ & $15.65\pm0.09$ & $\mathbf{6.5\pm0.5}$ & $6.4\pm0.1$ &
$5.6\pm0.3$ \\
NGC\,6441$^\dagger$ & $0.12\pm0.07$ & 
$16.35\pm0.07$ & $\mathbf{13.1\pm 0.7}$ & 
$0.54\pm 0.07$ & $16.93\pm0.07$ & $\mathbf{12.7\pm1.0}$ & $12.7\pm0.2$ &
$12.7\pm1.8$ \\
NGC\,6626 & $0.16\pm0.10$ & 
$14.47\pm0.11$ & $\mathbf{5.6\pm 0.3}$ & 
$0.59\pm0.10$ & $15.22\pm0.12$ & $\mathbf{5.9\pm0.5}$ & $5.4\pm0.1$ &
$5.1\pm0.3$ \\
NGC\,6638 & $0.24\pm0.09$ & 
$15.64\pm0.10$ & $\mathbf{9.6\pm 0.5}$ & 
$0.67 \pm 0.09$ & --- & --- & $9.8\pm0.3$ & $9.0\pm1.0$ \\
NGC\,6642 & $0.18\pm0.10$ & 
$15.25\pm0.11$ & $\mathbf{8.2\pm 0.7}$ & 
$0.61 \pm 0.10$ & --- & --- & $8.0\pm0.2$ & $8.3\pm0.9$ \\
NGC\,6717 & $0.16\pm0.11$ & 
$14.72\pm0.15$ & $\mathbf{7.3\pm 0.5}$ & 
$0.59\pm0.11$ & $15.11\pm0.15$ & $\mathbf{7.5\pm0.6}$ & $7.5\pm0.1$ &
$8.9\pm0.9$ \\
\hline
\end{tabular}
\tablefoot{$^\dagger$ Helium-enhanced BaSTI ZAHB models were applied,
with $Y=0.350$, due to the different category of the RRLs (OoIII).}
\end{table*}

The second-order polynomial fits for the models with canonical He
with the uncertainties on the coefficients are as follows:
\begin{multline}
    \ M_V = (1.047 \pm 0.028) + (0.456 \pm 0.063)\cdot\mathrm{[Fe/H]} +\\
    (0.077\pm 0.030)\cdot\mathrm{[Fe/H]}^2
    \label{eq:1}
\end{multline}
\begin{multline}
    \  M_I = (0.619 \pm 0.028) + (0.455 \pm 0.063)\cdot{\rm [Fe/H]} +\\
    (0.075 \pm 0.030)\cdot\rm{[Fe/H]}^2 .
    \label{eq:2}
\end{multline}


The $M_V-\rm{[Fe/H]}$ relations from \citet{2017A&A...605A..79G} and \citet
{2018MNRAS.481.1195M} are also presented in the upper panel of Fig.~\ref
{fig:mFEH-plot}. Our derived relations from BaSTI ZAHB models are more
consistent with \citet{2017A&A...605A..79G} but they deviate
towards fainter magnitudes at higher metallicites.
For a metallicity of $\rm{[Fe/H]=-1.15}$, which is the average metallicity of
our sample and the metallicity of the prototypical RR Lyrae star, the \citet
{2018MNRAS.481.1195M} relation would result in a $0.12$ smaller $(m-M)_V$,
underestimating the final distances.

\subsection{Reddening laws and coefficients}
\label{sec:reddening}

For both the $V$ and $I$ filters, subtracting the average of the mean
magnitudes and the luminosity--metallicity relation evaluated for the assumed
$\rm{[Fe/H]}$ gives the apparent distance modulus, which in turn provides the
final distance with the proper reddening laws and coefficients. Our aim is to
obtain the most up-to-date photometric distances in the $I$ band. For this,
three main points on the reddening laws have to be examined: \textit{(i)} the
average reddening law, quantified as $R_V = A_V/E(B-V)$; \textit{(ii)} the
corresponding extinction coefficient for the $I$ filter, represented by $A_I/
E(B-V)$ or $A_I/A_V$; and \textit{(iii)} the dependence on the effective
temperature of the extinction \citep[see, e.g., the discussion in][]
{2009ApJ...697..965B}.
The effect of these
transformations is increasingly important for higher reddening.

As far as what concerns the first point, the total-to-selective extinction
ratio $R_V$ is
commonly assumed as the constant of correlation between extinction $A_V$ and
reddening $E(B-V)$, but it is not constant and depends on the intrinsic
colour and the amount of reddening \citep[e.g.][]{1956ApJ...123...64B,
1975PASP...87..349O}. The $R_V$ ratio has been obtained by several authors in
the past, from \citet{1958AJ.....63..201W} and \citet{1965ApJ...141..923J},
and more recently by \citet{1998ApJ...500..525S} and \citet
{2016ApJ...821...78S}, based on a large number of stars from the SDSS, 2MASS,
and Pan-STARRS surveys. A deep and clear review is given in \citet
{2004AJ....128.2144M}. While there is a variety of $R_V$ values for specific
regions
of the Galaxy, ranging from 2.8 up to 5.0 \citep[e.g.][which used early-type O
and B stars]{1989ApJ...345..245C,1999PASP..111...63F}, there is a convergence
to $\sim 3.1$. This value is very close to the earlier value of 3.0
\citep{1961AN....286..113S}.

Surprisingly, \citet{2016ApJ...821...78S} obtained an average $R_V$ of $3.32
\pm0.18$ from thousands of stars sparse in the Galactic plane (see their
Fig.~15). As explained by the authors, this relatively higher value (3.3 vs.
3.1) is justified by the fact that their sample is based on a wide sample of
unselected red giant stars with an average temperature of about 4500 K and
$E(B-V)=0.65$. This corresponds to $B-V=1.0-1.1$ (K stars).

As discussed in \citet{2014MNRAS.444..392C}, for a given $E(B-V)$, the $R_V$
ratio is almost a linear function of the spectral type since the reddening
produces a different effective wavelength for each filter. For MW stars
obscured by normal dust, \citet{schmidt1982landolt} related the effective
total-to-selective extinction ratio with the reddening with the following
equation: $R_V = R_V^{00} + 0.28(B-V)^0 + 0.04E(B-V)$, where $(B-V)^0$ is the
intrinsic colour, and $R_V^{00}$ is the $R_V$ ratio for a star of zero-colour
in the limit of zero reddening \citep{2004AJ....128.2144M}.

Correcting the \citet{2016ApJ...821...78S} value with the Schmidt-Kaler
equation to A0 (Vega) type stars, we obtained $R_V^{00}=3.01$, which is
coherent with the earlier references within the errors.
In our case of \textit{V} photometry \citep{2001AJ....122.2587C} for RR Lyrae
variables, we adopted an average $E(B-V)=0.5$ and $(B-V)=0.35$ (A5--F5 spectral
type). Adopting the zero-reddening extrapolation of $R_V^{00}=3.07$ for Vega
\citep{2004AJ....128.2144M}, we obtained a coefficient of $R_V=3.19$.
However, the fact that the present sample is located close to the Galactic
plane and towards the bulge suggests that $R_V$ may be lower than $3.1$, as
discussed in recent works \citep{2013ApJ...769...88N,2014MNRAS.444..392C,
2019ApJ...874...30S}. By combining optical and NIR data,
\citet{2021ApJ...913..137P} argue that $R_V=2.7$ in the direction of the
bulge GC NGC\,6440 ($\ell=7\fdeg73$, $b=+3\fdeg80$). For our sample GCs
($5^\circ< |b|< 10^\circ$), we adopted a more conservative value of
$R_V=3.0$.





The second point listed above concerns the wavelength dependence of the
interstellar extinction, more specifically in the $I$ band, for which the OGLE
data are available. One of the most updated references on $BVI$ extinction
relations is \citet{1999PASP..111...63F}, who provided the extinction curve for
several wavelengths and computed extinction ratios for the Johnson and
Str\"omgren filters. For the $I$ filter, \citet{1999PASP..111...63F} derived a
ratio of $A_I/E(B-V) = 1.57$, which is very close to the earlier value of
$1.50$ from \citet{1975A&A....43..133S}.

However, \citet{1999PASP..111...63F} derived those relations for very blue stars
($T_{\rm eff}=30\,000$\,K and $\log g=4.0$), while we used RR Lyrae stars with
temperatures just lower than $10\,000$\,K. For this temperature regime, the
ratio should be greater than $1.57$, resulting in smaller distances
for heavily reddened GCs. In following work, \citet{2004AJ....128.2144M}
obtained $A_I/E(B-V)=1.71$ (see their Table~1), under the
assumption of $R_V=3.07$ for Vega at zero-reddening extrapolation.

The sensitivity of this coefficient to the temperature of the stars and to the
reddening is not very high and can be obtained using PARSEC \citep[\textit
{PAdova and TRieste Stellar Evolution Code}\footnote{\url{http://stev.oapd.inaf.it/cgi-bin/cmd}};]
[]{2012MNRAS.427..127B}
isochrones fitted with the reddening law at different
$A_V$ values. The PARSEC isochrones are not fully reliable to provide the zero
point transformations, since they assume $R_V=3.1$ for G2V stars (this value
was derived for a Vega-type star and gives about $3.3$ when extrapolated to a
G2V star). Scaling the coefficient with PARSEC, for the same values $B-V=0.3$
and $A_V=1.5$, the derived correction from Vega colours is only 0.007, which
is much smaller than the analogous $A_V/E(B-V)$ correction of 0.12 and is an
advantage of adopting the \textit{I} band instead of $V$.

In conclusion, assuming the correct $E(B-V)$ is used, the correct coeffient
is $A_I/E(B-V)=1.703$ for the RR Lyrae stars. Applying the correction due to
the bulge region in $R_V$ from 3.19 to 3.0 (a factor of 0.94 smaller), we
obtained a final coefficient ratio of $A_I/E(B-V)=1.60$.

The last point to be examined is the reddening variations of the coefficients.
All the $E(B-V)$ input values from the literature (Sect.~\ref
{sec:LitValues}) have been obtained from SGB F-type stars at the HB level,
which corresponds to an intrinsic $(B-V)=0.8-0.9$. Therefore, the spectra
are very different from the Vega and RR Lyrae stars ($B-V=0.3$, on average),
and they
require the temperature-reddening correction. We can estimate this correction
by analysing two figures from \citet{2004AJ....128.2144M}: first transform
$B-V$ into $B-I$ using Fig.~4, and then using their Fig.~8. From this analysis,
we obtained a correction factor of $0.85/0.92$ or $1/1.07$. The literature
$E(B-V)$ should be decreased by about $1/1.07$ (or increased by a factor
$1.07$) to get the correct reddening for the RRLs.

\begin{figure*}
    \centering
    \includegraphics[width=0.44\textwidth]{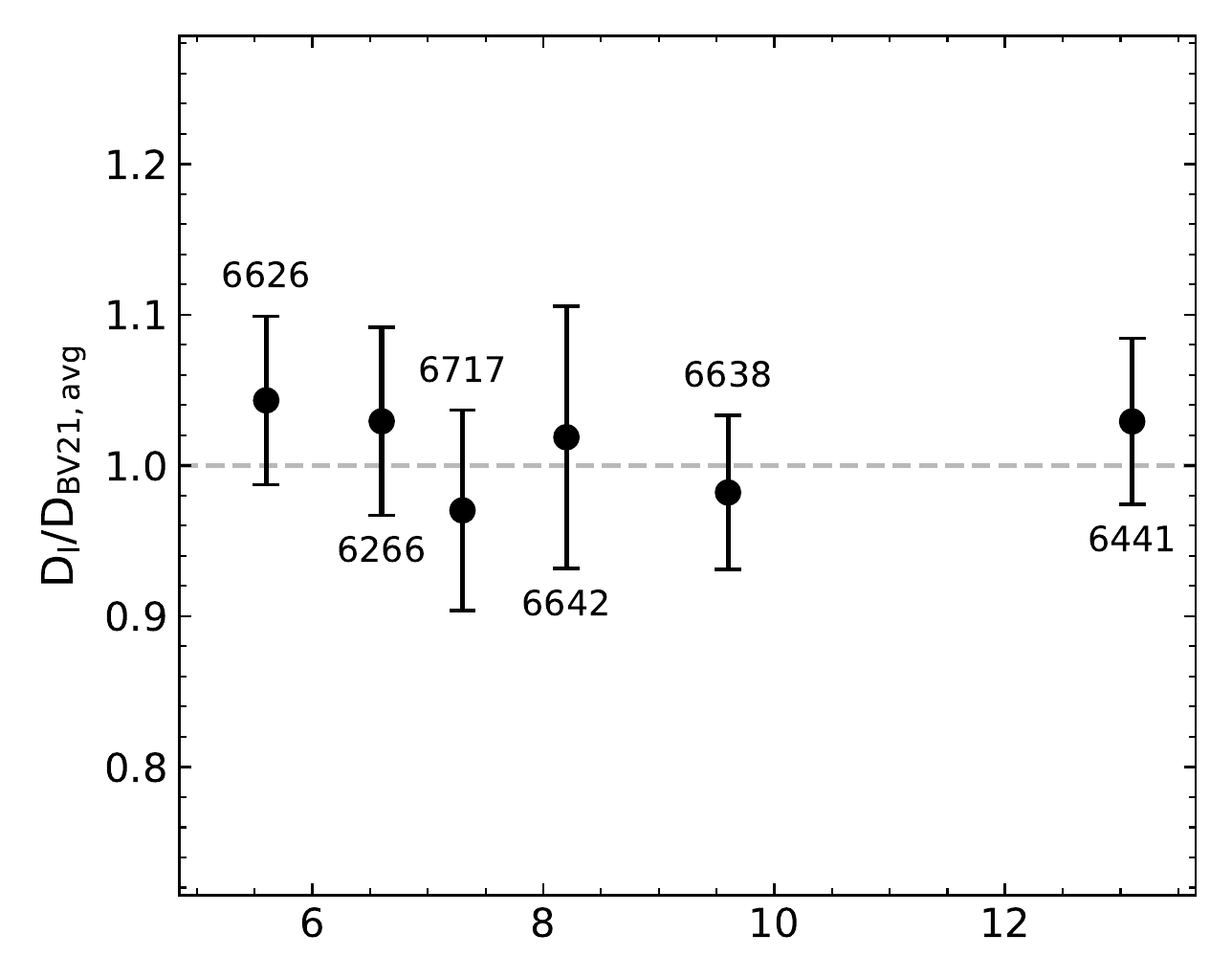}
    \hspace{2mm}
    \includegraphics[width=0.44\textwidth]{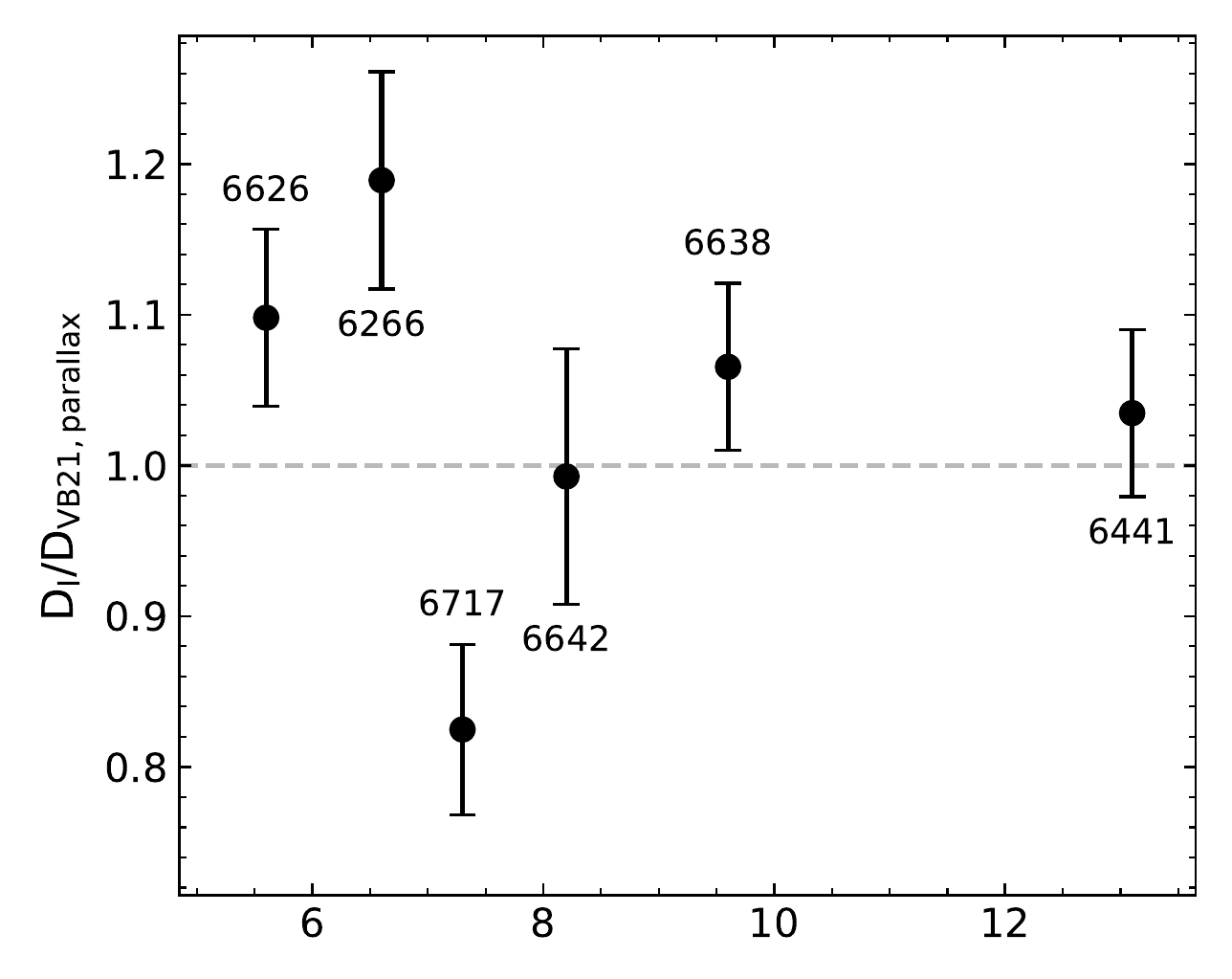}
    \caption{Ratio of the distances derived in this work ($D_I$, Table~\ref{tab:Results-I}), with the average distances from \citet[BV21,
    left panel]{2021MNRAS.505.5957B} and the distances derived from parallaxes
    from \citet[VB21, right panel]{2021MNRAS.505.5978V}. The vertical error
    bars correspond to the derived uncertainties, divided by D$_{\rm BV21}$ or
    D$_{\rm VB21}$. The distances from \citet{2021MNRAS.505.5978V} are
    discrepant with ours only for the three closest GCs (10\% for
    NGC\,6626, and 20\% for NGC\,6266 and NGC\,6717).}
    \label{fig:comp_dist}
\end{figure*}

Gathering these corrections on the extinction coefficients and reddening, the
equations we adopted to obtain the extinctions in the $V$ and $I$ filters
are as follows:
\begin{equation}
    \ A_V = [E(B-V)\cdot 1.07]\cdot 3.0 
    \label{eq:4}
\end{equation}
\begin{equation}
    \ A_I = [E(B-V) \cdot 1.07] \cdot 1.6. 
    \label{eq:5}
\end{equation}

In the three clusters with available data in \textit{VI} bands, it is possible
to calculate an $E(V-I)$ value for each RRL by subtracting the $(V-I)_*$
colour from the apparent magnitudes (Sect.~\ref{subsec:average}) from the
absolute $(M_V-M_I)$ colour (Sect.~\ref{subsec:basti}). The result for each RRL
could also account for the differential reddening across the cluster. The
colour excess $E(V-I)$ is also defined as $A_V-A_I$, and it can be converted to
$E(B-V)$ with Eqs.~\eqref{eq:4} and \eqref{eq:5}:
\begin{equation}
    \ E(V-I) = [E(B-V)\cdot 1.07] \cdot (3.0-1.6)
\end{equation}
\begin{equation}
    \ E(B-V) = \frac{(V-I)_* - (M_V-M_I)}{1.4 \cdot 1.07}.
\end{equation}

The resulting average $E(B-V)$ values are coherent with the literature ones
(Sect.~\ref{sec:LitValues}), with a small dispersion among the cluster RRLs.
Given that the computed reddening did not vary much, becoming irrelevant to
detect differential reddening, we opted to just update the input $E(B-V)$ with
an average of the computed $E(B-V)$ for these three GCs, and not to use it
separately star-by-star. Therefore, we updated the inputs as follows: $0.47\pm
0.05$ to $0.50\pm0.05$ for NGC\,6266, $0.47\pm0.05$ to $0.44\pm0.05$ for
NGC\,6441, and $0.43\pm0.04$ to $0.42\pm0.04$ for NGC\,6626.


%


\section{Results on the final distances}
\label{sec:results-dist}

All the methods and equations described in Sect.~\ref{sec:methods}, along with
a complete uncertainty propagation, were applied to obtain the final distances.
Table~\ref{tab:Results-I} shows the results of the absolute magnitudes,
distance moduli, and distances using both $I$ and $V$ bands. For NGC\,6638 and
NGC\,6642, we could not derive the distance with the mean $V$ magnitudes
since they are not available in \citet{2001AJ....122.2587C} catalogues.

Despite the study of the mean $I$ magnitudes of the RRLs with the new
OGLE-IV data \citep{2019AcA....69..321S} being our main science case, the
calculations with the $V$ magnitudes \citep{2001AJ....122.2587C} contribute
to the argument about the systematics between the different bands. From
Table~\ref
{tab:Results-I}, it is clear that the results with $\left \langle V \right
\rangle$, instead of being close to the ones with $\left \langle I \right
\rangle$, have higher uncertainties,
which may be explained by the differential reddening, lower
statistics of RRLs, and an inhomogeneity of the compiled data from \citet
{2001AJ....122.2587C} as compared to OGLE data.

Figure~\ref{fig:comp_dist} compares the ratio of our distances derived with
the \textit{I} band (D$_{\rm I}$) with those from \citet{2021MNRAS.505.5957B}
and \citet{2021MNRAS.505.5978V} listed in Table~\ref{tab:Results-I},
as a function of D$_{\rm I}$.
The difference between our distances D$_{\rm I}$ and the literature average
from \citet{2021MNRAS.505.5957B}
is less than 5\% for the six GCs and within our derived uncertainties.
On the other hand, the comparison with \citet{2021MNRAS.505.5978V} shows a
discrepancy up to $10-20$\% for the three closest clusters (NGC\,6626,
NGC\,6266, and NGC\,6717). The uncertainties of $\sim10\%$ in \textit{Gaia}
EDR3 parallaxes, the high-reddening region of the bulge, and a possible offset
of 0.007\,mas in the zero-point correction from \citet{2021A&A...649A...4L}
may explain this discrepancy.




The two sample clusters with fewer RRLs (NGC\,6642 and NGC\,6717) provided
interesting results. For NGC\,6717, the only RR Lyrae, with $\langle B\rangle
= 15.75$ \citep{1979SvA....23..284G} or $\langle V\rangle = 15.70$ \citep
{1999A&AS..136..237O}, is exactly the same one that is the only cluster RRL
member, and that sets the $\langle I \rangle$ average around 14.88
(Figs.~\ref{Fig:pmMemb} and \ref{fig:AvgMags2}).
The derived D$_V$ and D$_I$ distances are very close to \citet
{2020ApJ...891...37O}, and to the average from \citet{2021MNRAS.505.5957B}
which is mainly based on optical models fitting for this cluster. For
NGC\,6642, the reddening values are divergent. It seems that the $E(B-V)=0.6$
from \citet{2007AJ....133.1287V} is overestimated, and would result in a lower
distance. A new analysis with deep photometry and isochrone fitting
would be an important validation test for this cluster.

\section{Conclusions}
\label{sec:conclusions}
The sample of six globular clusters was selected for having newly identified
lists of RR Lyrae from the OGLE survey. It is interesting to note that  \citet
{2020MNRAS.491.3251P} assigned four of these clusters to a bulge population,
and two of them (NGC\,6441 and NGC\,6638) as belonging to a thick disk
population. \citet{2019A&A...630L...4M} assigned all of them, except
NGC\,6441, as belonging to the main bulge, and NGC\,6441 would be in an
unassociated low-energy category.

The distances derived are compatible with those derived by \citet
{2019MNRAS.482.5138B} and the average distances from \citet
{2021MNRAS.505.5957B}, considering the uncertainties.
When comparing them to the distances derived from \textit{Gaia} EDR3 parallaxes
in \citet{2021MNRAS.505.5978V}, a discrepancy of $10-20$\% is observed for
NGC\,6266, NGC\,6626, and NGC\,6717, which can be explained by some limitations,
possibly an offset in the zero-point correction from \citet{2021A&A...649A...4L}.
Since our method is based on a bona fide sample of member RRLs,
recent BaSTI models, correct relations for reddening, and a robust method, it was
already expected that our distances would be very close to the average distances from \citet{2021MNRAS.505.5957B}, based mainly on photometric derivations.

A wider sample will be adopted in a future work in order to revise
the bulge cluster distance, related to the Sun-Galactic centre distance \citep
{2016ARA&A..54..529B}. In fact, the uncertainties of $5-8\%$ obtained in this method show it is a
solid basis for new distance measurements, as a complement to the recent
geometric methods. The final catalogues of RRLs with data from \citet
{2001AJ....122.2587C}, OGLE-IV \citep{2019AcA....69..321S} and \textit{Gaia}
EDR3 \citep{2021A&A...649A...1G}, and the computed membership values are provided
in \texttt{VizieR}.


\begin{acknowledgements}
R.A.P.O. and S.O.S. acknowledge the FAPESP PhD fellowships nos. 2018/22181-0
and 2018/22044-3 respectively. B.B., L.O.K. and E.B. acknowledge partial
financial support from FAPESP, CNPq, and CAPES - Finance Code 001. S.O.
acknowledge partial support by the Universit\`a degli Studi di Padova Progetto
di Ateneo BIRD178590. S.C. acknowledges support from Progetto Mainstream INAF
(PI: S. Cassisi), from INFN (Iniziativa specifica TAsP), and from PLATO
ASI-INAF agreement n.2015-019-R.1-2018. APV acknowledges the DGAPA-PAPIIT
grant IG100319.
This work has made use of data from the European Space Agency (ESA) mission
{\it Gaia} (\url{https://www.cosmos.esa.int/gaia}), processed by the {\it Gaia}
Data Processing and Analysis Consortium (DPAC,
\url{https://www.cosmos.esa.int/web/gaia/dpac/consortium}). Funding for the DPAC
has been provided by national institutions, in particular the institutions
participating in the {\it Gaia} Multilateral Agreement.
We thank an anonymous referee for the remarks that improved this paper.
\end{acknowledgements}

%
\bibliographystyle{aa} 
\bibliography{Paper-OGLE}
%

\end{document}